\magnification \magstep1
\input amssym.def
\input amssym.tex
\bigskip
\bigskip
\bigskip
\bigskip
\bigskip
\centerline{\bf Scalar--Field Amplitudes in Black--Hole Evaporation}
\bigskip
\centerline{A.N.St.J.Farley and P.D.D'Eath}
\bigskip
\centerline{Department of Applied Mathematics and Theoretical Physics,
Centre for Mathematical Sciences,} 
\smallskip
\centerline{University of Cambridge, Clarkson Road, Cambridge CB3 0WA,
United Kingdom}
\bigskip
\centerline{Abstract}
\bigskip
We consider the quantum-mechanical decay of a Schwarzschild-like black
hole into almost-flat space and weak radiation at a very late time.
That is, we are concerned with evaluating quantum amplitudes (not just 
probabilities) for transitions from initial to final states. In this 
quantum description, no information is lost because of the black hole.
The Lagrangian is taken, in the first instance, to consist of the
simplest locally supersymmetric generalization of Einstein  gravity
and a massless scalar field.  The quantum amplitude to go from given 
initial to final bosonic data in a slightly complexified time-interval 
$T={\tau}{\,}{\exp}(-i{\theta})$ at infinity may be approximated by the 
form const.$\exp(-I)$, where $I$ is the (complex) Euclidean action of 
the classical solution filling in between the boundary data.  
Additionally, in a pure supergravity theory, the
amplitude const.$\exp(-I)$ is exact.  Suppose that Dirichlet boundary
data for gravity and the scalar field are posed on an initial
spacelike hypersurface extending to spatial infinity, just prior to 
collapse, and on a corresponding final spacelike surface, sufficiently 
far to the future of the initial surface to catch all the Hawking 
radiation.  Only in an averaged sense will this radiation have an 
approximately spherically-symmetric distribution.  If the
time-interval $T$ had been taken to be exactly real, then the 
resulting `hyperbolic Dirichlet boundary-value problem' would, as is 
well known, not be well posed.  Provided instead (`Euclidean
strategy') that one takes $T$ complex, as above 
($0<{\theta}{\;}{\leq}{\;}{\,}{\pi}/2$), one expects that the field
equations become strongly elliptic, and that there exists a unique 
solution to the classical boundary-value problem.  Within this
context, by expanding the bosonic part of the action to quadratic
order in perturbations about the classical solution, one obtains the 
quantum amplitude for weak-field final configurations, up to 
normalization.  Such amplitudes are here calculated for weak final 
scalar fields.\par
\bigskip
\bigskip
\parindent = 25pt
{\bf 1. Introduction}
\bigskip
\indent
Since 1976, the most generally accepted option for the end-point of
gravitational collapse to a black hole was that quantum coherence or
information would be lost [1].  The gratifying change in opinion this
past week [2] now makes it possible at last to begin publishing our
late-1990's work on another option, namely, that information is not in
fact lost, and that the end state is a combination of outgoing
radiation states (see option 3 below).  The present work began as a
doctoral dissertation at Cambridge of one of the present authors [3]
in 1997-2001 (the thesis was approved in summer 2002).  This brief
letter is an introductory sketch for numerous further papers contained
in the thesis, which will provide detailed derivations of quantum
amplitudes in black-hole evaporation.\par  
\smallskip
\indent
For simplicity, we fix attention in this paper on non-rotating black 
holes.  In the semi-classical theory, a black hole of initial mass 
$M_{0}$ slowly radiates away its mass into predominantly massless 
particles, over a timescale ${\sim}{\;}(M_{0})^3$ in Planck units 
[4,5].  Remarkably, at late retarded times, the emission depends only 
on the late-time quasi-stationary state of the `black hole' and not on 
the precise details of its formation.  The frequency spectrum has 
precisely the thermal character expected, with the Hawking temperature 
at infinity (units restored) given by 
$$T_{H}{\;}={{\;}{\hbar}{c^3}\over 8{\pi}GM_{0}{k_B}}{\;},\eqno(1)$$
where $k_B$ denotes Boltzmann's constant.\par
\smallskip
\indent
If one allows for the back-reaction of the emitted particles on the
gravitational field, then, to a good approximation, one can model
the black-hole evaporation with a sequence of quasi-stationary
Schwarzschild solutions until the mass has nearly reached the Planck
mass [4,5].  The Schwarzschild mass-loss rate balances the rate at
which energy is radiated to infinity.\par
\smallskip
\indent
For the (ill-understood) very late stages of this evaporation, there
are a number of conceivable options [1,6-10].  Some of the main
options are:\par
\smallskip
\indent
1.  The gravitational collapse could leave behind a naked singularity
with negative mass (as there is no event horizon) which persists, 
thereby violating cosmic censorship;\par 
\smallskip
\indent
2.  A stable Planck-mass object might remain, containing the
negative-energy quanta and stellar matter, so that the joint quantum
state of the remnant and the emitted quanta would be pure;\par
\smallskip
\indent
3.  The object might disappear completely, but all information about
black-hole configurations and any locally conserved quantities might
still escape to infinity.  That is, particle correlations would be
restored, so that the final state of the quantum field would be
pure;\par 
\smallskip
\indent
4.  The object might disappear completely, taking with it the
information about the stellar matter, negative-energy quanta and any
quantities not coupled to long-range fields, so that the final state
would be mixed.  The nature of the final radiation in this instance,
and in option 3 above, is still to be uncovered.\par
\smallskip
\indent
This paper focusses on options 3 and 4 and probabilities or quantum 
amplitudes associated with such outcomes.  Below, we will describe 
how the total amplitude of (probable) weak-field final configurations 
can be computed from the imaginary part of the Lorentzian action, 
which is quadratic in the field perturbations in the linearised
theory.  The computation of such amplitudes for late-time scalar 
perturbations is the main result to be reported in this paper.\par
\smallskip
\indent
For simplicity, we take the bosonic matter to be described by a 
massless scalar field $\phi.$  The full Lagrangian might, in principle, 
be more complicated, containing also appropriate fermionic fields and 
(say) Maxwell or Yang-Mills fields. Indeed, although we shall not
study fermionic fields in this paper, it will be simplest (see Sec. 2)
to assume that the Lagrangian is that of [ ], consisting of
supermatter -- a complex massless scalar field and spin - 1/2 partner
-- coupled to $N=1$ supergravity, consisting of Einstein gravity with
gravitino partner.  In a linearised approximation, 
we expand the massless scalar field and the metric around a background
classical 
field configuration, in which $\phi$ is taken to be real, 
describing a (bosonic) spherically symmetric 
`matter' collapse, in which only the scalar field acts as a matter 
source for gravity.  For both scalar and gravitational fields,
`Dirichlet boundary data' (see below) are specified on a (edgeless)
spacelike hypersurface ${\,}{\Sigma}_i{\,}$ at an early time -- 
roughly the moment of collapse -- and on a final late-time spacelike 
hypersurface ${\,}{\Sigma}_f{\,}$, long after the black hole has 
evaporated.  The location of ${\,}{\Sigma}_f{\,}$ will be such as to 
register all the evaporated radiation.  Both surfaces extend to
spatial infinity.\par
\smallskip
\indent
To fix one's physical intuition, it may be imagined that the energy
density of the initial scalar configuration is extremely diffuse, such
that almost all of the mass is distributed over radii vastly greater
than the `Schwarzschild radius' $2M_{0}{\,}$.  Further, the initial
time-variation of the scalar and the gravitational fields
can be taken to be extremely slow.\par
\smallskip
\indent
Away from the black hole, typical wavelengths of perturbation modes
are small in comparison with the local radius of curvature of the
space-time -- the characteristic length scale over which the background
space-time changes significantly.  Prior to the black hole's final
disappearance, however, typical wavelengths of perturbation modes
which will contribute significantly to radiation in the late-time
quantum treatment are comparable in magnitude to the black-hole
radius.  Correspondingly, one finds that the radiating
Schwarzschild-like Vaidya metric [11-13] models, to a good
approximation, the background space-time of an evaporating black hole 
at late times, outside the central strong-field region.  Such a 
semi-classical description will need to be joined onto a comparable 
treatment of the more non-linear central regions.\par
\smallskip
\indent
Nevertheless, there is an intrinsically `non-classical' character to this
formulation, since wave-like boundary-value problems are well known to be
badly posed [14].  Analogously, the Uncertainty Principle forbids 
the simultaneous specification of a dynamical variable and its 
conjugate momentum -- precisely the data specified in a Cauchy
problem.  The price that we pay for our boundary-value formulation 
may be a lack of existence and/or uniqueness in the solution of the 
boundary-value problem, in the case of a real time-interval $T$ at 
infinity; this is intimately connected with the occurrence of 
singularities.  But experience of other simpler such `hyperbolic 
boundary-value problems', such as the flat-space wave equation, 
leads one to expect that there will be a (complex) classical
solution, typically unique, provided that we rotate the proper-time 
separation at infinity, $T$, infinitesimally into the complex, 
{\it \`a la} Feynman, and indeed that a real Riemannian solution will
be reached when $T$ is rotated by a phase of $\pi/2$ to a Euclidean
proper distance $\tau = iT.$    There is a substantial mathematical theory of 
such {\it strongly elliptic} Dirichlet boundary-value problems in the 
complex, with existence, uniqueness and regularity properties 
corresponding to the more familiar real elliptic case [15].\par
\bigskip
\indent
{\bf 1.1 Existence for the full Einstein--scalar system} 
\bigskip
\indent
A variety of spherically symmetric boundary--value examples in the 
complexified version of general relativity, coupled to a massless 
scalar field, are being worked out, and will then be published.  
One would like to show that (three-dimensional) asymptotically flat
initial and final boundary data (Dirichlet data) for the
spherically-symmetric gravitational and scalar fields can be
specified, together with (in the first instance) a purely Euclidean
'time-interval' ${\mid}T{\mid}$, such that there is a smooth classical
Riemannian infilling solution of the coupled Einstein/massless scalar field
equations, agreeing with the given Dirichlet boundary data.  This
requirement is
perhaps not unreasonable, since the coupled Riemannian field equations 
are 'elliptic {\it modulo} gauge', and elliptic equations tend to
smooth out boundary data.  Further, that,  as the
time-interval-at-infinity $T$ is rotated back in the complex, towards 
the Lorentzian region, by a phase less than ${\pi}/2$, while
holding fixed the Dirichlet data on the initial and final surfaces,
then the full complex classical solution varies smoothly with the
phase, as does the classical action. \par
\smallskip
\indent
In particular, if the scalar-field
part of the Dirichlet data is chosen to be sufficiently strong, then
the corresponding Lorentzian-signature solution, reached by a phase
rotation by ${\pi}/2$, will describe collapse to a black hole.
Only as one approaches a phase of ${\pi}/2$ would the classical 
solution become singular.  Of course, this would be in accord with the 
singularity theorems of general Lorentzian-signature relativity 
[16].  This view of strongly elliptic {\it versus} singular 
Lorentzian-signature solutions also shows further the insight of 
Feynman's $+i{\epsilon}$ prescription.    
\par  
\smallskip
\indent 
Note that the mass $M$ of the classical 'space-time' is a functional
of the boundary data, including the possibly complex time-at-infinity
$T$ [17,18].  It has been found in the above examples that $M$ is
negative when $T = -i\tau$, for $\tau$ real, corresponding to a real
Riemannian solution.  The geometry is regular at the origin $r = 0$ of
spherical symmetry (and elsewhere) and the Riemannian solution lives
on a portion of Euclidean space $\Bbb R^4$.  One does not encounter
the different topology of the Riemannian positive-mass Schwarzschild
solution [19,20].  \par
\smallskip
\indent
As described in [17,18,21], each of the intrinsic spatial metrics
$h_{ijI}$ and $h_{ijF}$ on $\Sigma_I$ and $\Sigma_F$ will have an
intrinsic Arnowitt-Deser-Misner mass $M_{ADM}$ [22,23], measured by the
usual $1/r$ fall-off rate of the spatial metric near spatial
infinity. But (see Sec.4.4 of [18]) in general, for the classical
infilling metric $g_{\mu\nu}$, the obvious surfaces of constant time
will be emitted in such a way that the mass $M$ of the 4-dimensional
geometry differs from the '3-dimensional mass' $M_{ADM}$.  It is this
which allows for the possibility that the mass $M$ of the classical
geometry may depend on the time-separation $T$ at infinity, as well as
on the bounding 3-metrics $h_{ijI}, h_{ijF}$, as in the previous
paragraph. \par
\smallskip
\indent
Note also:  it is essential that the 3-metrics prescribed on
$\Sigma_I$ and $\Sigma_F$ should each have the same intrinsic ADM mass
$M_{ADM}$.  Otherwise [17,18], any classical infilling 'space-time' will
have these 3-surfaces badly embedded near spatial infinity. \par
\smallskip
\indent
In the more general non-spherical black-hole problem to be studied in
the rest of this paper, the slight complexification of $T$ induces an 
imaginary part in the total (Lorentzian) action, which leads (below)
to the quantum amplitudes and probabilities needed to examine options 3 
and 4 above.\par
\bigskip
\indent
{\bf 2. The quantum amplitude}\par
\bigskip
\indent
Consider first quantum field theory for gravity coupled to bosonic and
fermionic fields. In particular, take the case in which non-trivial
bosonic data are specified on the boundary surfaces, while for
simplicity the fermionic boundary data are taken to be zero.
The `Euclidean' quantum amplitude to go between the prescribed initial and
final data is given formally by a Feynman path integral. Naively,
disregarding the problem of infinities, one would expect the amplitude
to have the asymptotic form
$${\rm Amp}{\;}{\;}{\sim}{\;}{\;}({A_0}+{\hbar}{A_1}+{{\hbar}^2}{A_2}+...)
{\;}{\exp}(-{I_B}/{\hbar}){\;}.\eqno(2)$$
Here $I_B$ is the real classical `Euclidean' action of a Riemannian
solution of the coupled Einstein/bosonic-matter classical field
equations.  [For simplicity, we assume that there is a unique
classical solution.]  Further, $I_B$ and the loop terms
${A_0},{A_1},{A_2},...$ depend in principle on the boundary data.  
For our case of matter coupled to Einstein gravity, 
${I_B},{A_0},{A_1},...$ also obey differential constraints connected 
with the local coordinate invariance of the theory, and with any 
other local invariances such as gauge invariance [24]. \par
\smallskip
\indent
In particular, when the theory is also invariant under local 
supersymmetry, and when the fermionic boundary data are now allowed to
be non-zero, the above semi-classical expansion (2) becomes extremely
simple.  For example, for $N=1$ supergravity, one has [18,25]: 
$${\rm Amp}{\quad}{\sim}{\quad}{A_0}{\;}{\exp}(-I/{\hbar}){\;}.\eqno(3)$$
This is not just a formal expression, as in Eq.(2), but an exact
statement, freed of infinities. The 'one-loop factor' $A_0$ is in fact
a constant, while $I$ 
denotes the action of the classical solution, which includes both
bosonic and fermionic parts.  Thus, in this case, the
classical action is all that is needed for the quantum
computation.  A corresponding situation arises with ultra-high-energy
collisions, whether between black holes [26], in particle scattering
[27], or in string theory [28]. For more complicated field theories
possessing local supersymmetry, such as supergravity coupled to
supermatter [29,30], quantum amplitudes may well still be meaningful
since free of divergences.  Their semi-classical expansion is,
however, expected to have the more general form (2).  \par    
\smallskip
\indent
The difficulty of assigning meaning to quantum amplitudes in field
theories which contain Einstein gravity and other lower-spin fields,
but which are not invariant under local supersymmetry, might also
encourage one to turn to string theory.  But it is again true in
string theory that only for the superstring models, which are
invariant under local supersymmetry transformations, does one have
meaningful finite amplitudes [31,32].  \par
\smallskip
\indent
In the asymptotically-flat, spatially-${\Bbb R}^3$ context here, the
purely Riemannian case above corresponds to a time-separation at
spatial infinity of the usual rotated form $T=-i{\tau}$, where
${\tau}$ is a positive imaginary-time separation. Consider further the
locally supersymmetric case, but take only the bosonic data to be
non-zero (real) on the initial and final surfaces. Suppose that there
is a unique infilling Riemannian classical solution. Following the
standard route,  one can
study the (now complex) bosonic amplitude (2) or (3) as $T$ is rotated
through angles ${\theta}$ from ${\theta}={\pi}/2$ to
${\theta}=+{\epsilon}$, with
$$T{\;}={\;}{\exp}(-i{\theta}){\,}{\tau}{\;}.\eqno(4)$$
Provided that there continues to exist a (now complex) bosonic
classical solution to the Dirichlet problem, as ${\theta}{\,}>{\,}0$
decreases from ${\pi}/2$ towards zero, the expression (2) or (3) should
still continue to give the form of the (analytic) quantum amplitude.
In particular, this would occur if strong ellipticity held for the
coupled Einstein/bosonic-matter field equations.\par
\bigskip
\indent
{\bf 3. The classical action and amplitude for weak perturbations}
\bigskip
\parindent = 18pt
\indent
{\bf 3.1 The approximate space-time metric}\par
\medskip
\parindent = 25 pt
\indent
The classical background bosonic fields, the metric $g_{{\mu}{\nu}}$
and (real) scalar field ${\phi}$, are each taken to have a `large'
time-dependent 
spherically symmetric part and a `small' perturbative part, which can
be expanded out in terms of sums over tensor, vector and scalar
harmonics [33-39].  Each harmonic is weighted by a function of the 
time- and radial coordinates $(t,r)$.  The `large' or `background' Lorentzian 
space-time metric can be put in the form [40,41]:
$$ds^2{\;}
={\;}-e^{b(t,r)}{\,}dt^2{\,}+{\,}e^{a(t,r)}{\,}dr^2{\,}
+{\,}r^2{\,}(d{\theta}^2{\,}
+{\,}{{\sin}^2}{\theta}{\,}d{{\varphi}^2}){\;}.\eqno(5)$$
The `background' scalar field is denoted by ${\Phi}(t,r).$  The classical
spherically-symmetric part of the Einstein and scalar field equations
will be as given in [40] (`Lorentzian') or [41] (Riemannian) for an 
exactly spherically-symmetric set of fields, except for an additional 
effective energy-momentum contribution $<T^{EFF}_{{\mu}{\nu}}>$, 
resulting from local space-time averaging of the contribution to the 
Einstein equations of terms quadratic and higher in the combined 
perturbations [42,43]. It will be assumed that 
$<T^{EFF}_{{\mu}{\nu}}>$ is defined in such a way as to be spherically 
symmetric.\par
\bigskip
\parindent = 18pt
\indent
{\bf 3.2 Scalar field -- harmonic decomposition}\par
\medskip
\parindent = 25 pt
\indent
Now consider purely bosonic weak classical perturbations around this
dynamic background.  For simplicity of exposition, consider a `small'
real scalar perturbation, denoted by ${\phi}^{(1)},$ where 
${\phi}{\,}={\,}{\Phi}{\,}+{\,}{\phi}^{(1)}.$   Because of the 
approximate spherical symmetry, one may expand the perturbation 
${\phi}^{(1)}$ in the form
$${\phi}^{(1)}(t,r,{\theta},{\varphi}){\;}
={\;}{1\over r}{\;}{{\sum}_{{\ell}=0}^{\infty}}{\;}
{{\sum}_{m=-{\ell}}^{\ell}}{\;}
{Y_{{\ell}m}}({\Omega}){\;}{R_{{\ell}m}}(t,r){\;}.\eqno(6)$$  
Here, $Y_{{\ell}m}({\Omega})$ denotes the $({\ell},m)$ spherical 
harmonic of [44].  The boundary conditions on the radial functions 
$R_{{\ell}m}(t,r)$ as $r{\,}{\rightarrow}{\,}0$ 
follow from the regularity at $r{\,}={\,}0$.  Recall that, in the 
strongly elliptic case, all fields are analytic in the interior of 
the large cylindrical boundary formed by the initial and final 
surfaces together with a surface at large $r$.\par
\smallskip
\indent
The perturbed scalar wave equation at late times,
$${\nabla}^{\mu}{\,}{\nabla}_{\mu}{\,}{\phi}^{(1)}{\;}
={\;}0{\;},\eqno(7)$$
with respect to the background geometry, leads to the $({\ell},m)$ 
mode equation
$${{\bigl[}e^{(b-a)/2}{{\partial}_r}{\bigr]}^2}{R_{{\ell}m}}
-({{\partial}_t})^2{R_{{\ell}m}}
-{\biggl(}{1\over 2}{\biggr)}{\bigl[}{{\partial}_t}(a-b){\bigr]}
{\;}({{\partial}_t}{R_{{\ell}m}}) 
-{V_{\ell}}(t,r){\;}{R_{{\ell}m}}{\;}={\;}0{\;},\eqno(8)$$
where
$${V_{\ell}}(t,r){\;}
={\;}{{e^{b(t,r)}}\over {r^2}}{\;}
{\biggl[}{\ell}({\ell}+1){\,}+{\,}{{2m(t,r)}\over r}{\biggr]}{\;}\eqno(9)$$
is real and positive in the `Lorentzian' case.  Here, $m(t,r)$ is
defined by ${\exp}[-a(t,r)]=1-[2m(t,r)]/r{\;}.$  This potential
$V_{\ell}(t,r)$ generalises the well-known massless-scalar effective 
potential in the exact Schwarzschild space-time [23,39], which
vanishes at the event horizon ${\lbrace}r=2M{\rbrace}$ and at
infinity, and has a peak near ${\lbrace}r=3M{\rbrace}$.\par
\smallskip
\indent
There is, of course, an analogous harmonic decomposition of the weak
gravitational-wave perturbations about the
nearly-spherically-symmetric background [33-39].\par
\bigskip
\parindent = 18pt
\indent
{\bf 3.3 The classical action}\par
\medskip
\parindent = 25pt
\indent
After detailed calculation [3,45], one finds that the classical 
Lorentzian action for a classical bosonic solution to our model, 
including the quadratic-order contribution of weak perturbations, 
can be written as
$$S_{class}[h^{(1)}_{ij},{\phi}^{(1)}]{\;}={\;}{1\over {32{\pi}}}
{\int}_{{\Sigma}_f}{\,}d^3 x{\,}{\pi}^{(1)ij}{\,}h^{(1)}_{ij}{\;}+{\;}
{1\over 2}{\int}_{{\Sigma}_f}{\,}d^3x{\;}{{\phi}^{(1)}}{\,}
{{\pi}_{{\phi}^{(1)}}}{\;}
-{\;}MT{\;}.\eqno(10)$$
Here, for simplicity, it is assumed that the only non-zero
(prescribed) boundary perturbations are in the intrinsic spatial metric 
$h^{(1)}_{ij}$ and in the real part of the scalar field ${\phi}^{(1)}$
on the final  
surface ${{\Sigma}_f}{\,}$.  The corresponding perturbations in the
initial boundary data $h^{(1)}_{ij}$ and ${\phi}^{(1)}$ on
${\Sigma}_{i}$ have been taken to be zero.  Of course, one could
easily include them also.  Here ${\pi}^{(1)ij}$ is the linearized
perturbation of the momentum canonically conjugate to the intrinsic
spatial $g_{ij}$, given in [23].  Also,
$${\pi}_{{\phi}^{(1)}}{\;}
={\;}n^{(0){\mu}}{\,}{{\nabla}_{\mu}}{\phi}^{(1)}{\;}\eqno(11)$$
is the momentum canonically conjugate to ${\phi}^{(1)}$.  
Here, $n^{(0){\mu}}$ is the future-directed unit time-like normal 
vector to the three-surface (here ${\Sigma}_f$), in the `Lorentzian' 
case.
Further, in Eq.(10), $M$ is the '4-dimensional mass' of the 
`space-time', as in Sec.1.1.  Typically, $M$ will differ from the ADM
mass $M_{ADM}$ found from the intrinsic 3-metrics $h_{ijI}, h_{ijF}$.\par
\bigskip
\parindent = 18pt
\indent
{\bf 3.4 The quantum amplitude for scalar perturbations on the final
surface}\par
\medskip
\parindent = 25pt
\indent
On the initial surface ${\Sigma}_i{\,}$, we take the initial data such
that there is a quasi-stationary Schwarzschild-like background.  On
or near the final surface ${\Sigma}_f{\,}$, long after evaporation, 
one again expects a nearly-Schwarzschild background; the
characteristic time-scale over which the background geometry varies 
is very much greater than the period of a typical wave.  In such
cases, the radial wave functions $R_{{\ell}m} (t,r)$ can be 
decomposed harmonically with respect to $t$, and can be conveniently 
normalised (below).  Hence, near ${\Sigma}_f$ say, one can write
$${\phi}^{(1)}{\;}={\;}{1\over r}{\;}
{\sum}_{{\ell}m}{\,}{\int}_{0}^{\infty}{\,}dk{\;}a_{k{\ell}m}{\;}
R_{k{\ell}m}(t,r){\,}Y_{{\ell}m}({\Omega}){\;},\eqno(12)$$
where the ${\lbrace}a_{k{\ell}m}{\rbrace}$ are real quantities.  
Spherical symmetry implies that, on ${\Sigma}_f{\,}$, 
$$R_{k{\ell}m}(t,r){\;}={\;}R_{k{\ell}}(r){\;}\eqno(13)$$
is independent of $m$, where $R_{k{\ell}}(r)$ is a real function.
Note [see below, Eq.(20)] that the allowed values of $k$ are discrete,
labelled by an integer $n{\,}={\,}0,1,2,...$, and are given by 
$k{\,}={\,}k_{n}{\,}={\,}n{\,}{\pi}/{\mid}{\,}T{\,}{\mid}.$\par
\smallskip
\indent
We describe briefly the normalisation of the $R_{k{\ell}}(r)$.
With the notation ${\tilde r}{\,}={\,}kr$, the boundary
conditions of regularity as ${\,}r{\,}{\rightarrow}{\,}0{\,}$ imply
$$R_{k{\ell}}(r){\quad}{\propto}{\quad}r{\,}j_{\ell}({\tilde r}){\quad}
{\sim}{\;}{\quad}{\hbox{const.}}{\;}{\tilde r}^{{\ell}+1}{\;},\eqno(14)$$
where $j_{\ell}({\tilde r})$ are (real) spherical Bessel functions
[46].  For a background geometry which is nearly Schwarzschild at
large radius, one also has
$$R_{k{\ell}}(r){\quad}{\sim}{\quad}z_{k{\ell}}{\;}e^{ikr_S^*}{\;}
+{\;}{\bar z}_{k{\ell}}{\;}e^{-ikr_S^*}{\;}\eqno(15)$$
as $r{\,}{\rightarrow}{\,}{\infty}{\,}$, where, for large $r{\,}$,
$$r_S^*{\;}{\;}{\sim}{\;}{\;}r{\;}+{\;}2M{\;}{\log}{\biggl(}
{\Bigl(}{r\over {2M}}{\Bigr)}-1{\biggr)}{\quad}\eqno(16)$$    
is the Regge-Wheeler `tortoise' coordinate [23,33].  Here the 
${\lbrace}z_{k{\ell}}{\rbrace}$ are dimensionless complex
coefficients which are determined by the regularity above at
$r{\;}={\;}0$, subject to suitable normalisation [3].  The 
$z_{k{\ell}}$ are related to the Bogoliubov coefficients [47,48], thus
making contact with the original formulation of black-hole evaporation
[4,5].  The detailed relations are described in the fuller works
[3,45].\par 
\smallskip
\indent
Given the above information in a neighbourhood of the final surface
${\Sigma}_f{\,}$, one can compute the classical scalar contribution 
to the Lorentzian action $S_{\rm class}[h_{ij}^{(1)},{\phi}^{(1)}]$ 
of Eq.(10), in the case that the time-interval 
${\,}T{\;}={\;}{\tau}e^{-i{\theta}}{\,}$ at infinity is rotated 
slightly into the complex.  This classical scalar contribution will 
in general be complex.  The Euclidean action is then defined {\it via}
$$-{\,}I_{\rm class}{\;}={\;}i{\,}S_{\rm class}{\;}.\eqno(17)$$
In the limit that ${\theta}{\,}{\rightarrow}{\,}0$ (while keeping
${\theta}$ strictly positive), one finds that the (bosonic) 
quantum amplitude has the form (say, in the case of a locally supersymmetric
Lagrangian):
$${\rm Amp}{\;}{\;}{\sim}{\;}{\;}{\hbox{const.}}{\;}
{\exp}(-{\,}I_{\rm class}){\;}\eqno(18)$$
apart from possible loop corrections [Eq.(2)] which will only come
into play at or above the Planck energy.  This amplitude  has a
(computable) oscillating part in each mode, multiplied by a
product of Gaussians
$${\mid}{\rm Amp}{\mid}{\;}{\;}{\propto}{\;}{\;}
{\exp}{\biggl(}-{{4{\pi}^3}\over {{\mid}T{\mid}^2}}{\;}
{\sum}_{n=1}^{\infty}{\;}{\sum}_{{\ell}=0}^{\infty}{\;}
{\sum}_{m=-{\ell}}^{\ell}{\;}n{\;}
{\mid}z_{n{\ell}}{\mid}^2{\;}{\mid}a_{n{\ell}m}{\mid}^2
{\,}{\biggr)}{\quad}\eqno(19)$$
in the coordinates ${\lbrace}a_{k{\ell}m}{\rbrace}$ of Eq.(12) for the
perturbed final scalar data.  Here, because of the large-but-finite
time interval ${\mid}T{\mid}$, the eigen-frequencies are of the form
$$k{\;}={\;}k_n{\;}={\;}{{n{\pi}}\over {\mid}T{\mid}}{\quad}.\eqno(20)$$
\par
\parindent = 18pt
\indent
{\bf 4. Comments and further work}\par
\medskip
\parindent = 25pt
\indent
By using a field-based amplitude in quantum gravity and by rotating
the time interval $T$ at infinity slightly into the complex:
$T{\;}{\rightarrow}{\;}{\tau}{\;}e^{-i{\theta}}$, one expects to have
a strongly elliptic classical boundary-value problem.  For weak wave
fields of different spin at the final late-time boundary 
${\Sigma}_f{\,}$, after the black-hole evaporation, one can compute 
the {\it quantum amplitudes} for different final configurations as
${\exp}(-I_{\rm class})$, up to an overall factor.  An explicit
example of this for the scalar-field perturbations of our model was 
given in Sec.3.  [Full details will be given in [45] and further papers.] 
Analogous `Gaussian' results for the spin-1 Maxwell field and for spin-2 
gravitational-wave perturbations are contained in [3] and will also be
submitted; similarly for the fermionic case.\par
\smallskip
\indent
The late-time radiation can be viewed either in a field or in a
particle representation, and has the usual interpretation in terms of
thermal Hawking radiation.\par
\smallskip
\indent
Work is also under way on numerical solution of the complexified 
boundary-value problem for the coupled Einstein-scalar field 
equations, with real data posed on the initial and final
three-surfaces, and a complex time-interval $T$ specified at spatial
infinity, as considered above.  This requires generalisation of the
original Riemannian Einstein/scalar numerical boundary-value work of  
[41].  This should help one to understand, for example, the nature 
of any burst of radiation which might originate in the late stages 
of complexified `collapse'.\par
\smallskip
\indent
It would also be very interesting if similar methods could be applied
in the case that the spatial bounding surfaces are not just two copies
of ${\Bbb R}^3$.\par
\medskip
\indent
{\bf Aknowledgements}\par
\indent
We are grateful to an unnamed referee for suggestions which led to
substantial improvements in the Letter.\par
\bigskip
\goodbreak
\parindent = 1 pt

{\bf References}
\medskip
\indent
[1] S.W.Hawking, {\it Phys. Rev. D} {\bf 14}, 2460 (1976).\par
\indent
[2] S.W.Hawking, communication, GR17 Conference, Dublin, 18-24 July (2004).
\par
\indent
[3] A.N.St.J.Farley, 'Quantum Amplitudes in Black-Hole Evaporation',           
Cambridge Ph.D. dissertation, approved 2002 (unpublished).\par
\indent
[4] S.W.Hawking, {\it Nature (London)} {\bf 248}, 30 (1974).\par
\indent
[5] S.W.Hawking, {\it Commun. Math. Phys.} {\bf 43}, 199 (1975).\par
\indent
[6] S.W.Hawking, {\it Commun. Math. Phys.} {\bf 87}, 395 (1982).\par
\indent
[7] S.W.Hawking, `Boundary Conditions of the Universe' in {\it
Astrophysical Cosmology}, Proceedings of the Study Week on Cosmology
and Fundamental Physics, eds. H.A.Br{\"u}ck, G.V.Coyne and
M.S.Longair. Pontificia Academiae Scientarium: Vatican City,
{\bf 48}, 563 (1982).\par
\indent
[8] R.M.Wald, in {\it Quantum Theory of Gravity}, ed. S.Christensen, 
(Adam Hilger, Bristol) 160 (1984).\par
\indent
[9] P.H{\'a}j{\'i}{\v c}ek and W.Israel, 
{\it Phys. Lett. A} {\bf 80}, 9 (1980).\par
\indent
[10] J.Bardeen, {\it Phys. Rev. Lett.} {\bf 46}, 382 (1981).\par
\indent
[11] A.Das, M.Fischler and M. Ro\v cek, {\it Phys.Lett.B} {\bf 69}, 186 (1977).
\par
\indent
[11] P.C.Vaidya, {\it Proc. Indian Acad. Sci.} {\bf A33}, 264 (1951).\par
\indent
[12] H.Stephani, D.Kramer, M.A.H. MacCallum, C.Hoenselaers and E.Herlt,
{\it Exact solutions of Einstein's field equations}, 2nd. edition,
(Cambridge University Press, Cambridge) (2003).\par      
\indent
[13] A.Krasi{\'n}ski, {\it Inhomogeneous Cosmological Models},
(Cambridge University Press, Cambridge) (1997).\par
\indent
[14] P.R.Garabedian, {\it Partial Differential Equations}, 
(Wiley, New York) (1964).\par
\indent
[15] W.McLean, {\it Strongly Elliptic Systems and Boundary
Integral Equations}, (Cambridge University Press, Cambridge) (2000).\par
\indent
[16] S.W.Hawking and G.F.R.Ellis, {\it The large scale structure of
space-time}, (Cambridge University Press, Cambridge) (1973).\par
\indent
[17] P.D.D'Eath, {\it Phys.Rev.} {\bf D24}, 811 (1981).\par
\indent
[18] P.D.D'Eath, {~}{\it Supersymmetric {~}Quantum {~}Cosmology}, 
{~}(Cambridge {~}University {~}Press, {~}Cambridge) (1996).\par
\indent
[19] J.B.Hartle and S.W.Hawking, {\it Phys.Rev.}{\bf D13}, 2188
(1976).\par
\indent
[20] S.W.Hawking, 'The path-integral approach to quantum gravity', in
{\it General Relativity. An Einstein Centenary Survey},
eds. S.W.Hawking and W.Israel (Cambridge University Press, Cambridge)
(1979).\par 
\indent
[21] J.A.Wheeler, 'Superspace and the Nature of Quantum
Geometrodynamics' p.303, in {\it Battelle Rencontres}, ed. C.M.DeWitt
and J.A.Wheeler (W.A.Benjamin, New York) (1968). \par
\indent
[22] R.Arnowitt, S.Deser and C.W.Misner, 'Dynamics of General
Relativity', in {\it Gravitation: An Introduction to Current Research},
ed. L.Witten (Wiley, New York) (1962).\par
\indent
[23] C.W.Misner, K.S.Thorne and J.A.Wheeler, {\it Gravitation},
(Freeman, San Francisco) (1973).\par
\indent
[24] P.A.M.Dirac, {\it Lectures on Quantum Mechanics}, (Academic
Press, New York) (1965). \par
\indent
[25] P.D.D'Eath, 'What local supersymmetry can do for quantum
cosmology', in {\it The Future of Theoretical Physics and Cosmology}, 
eds. G.W.Gibbons, E.P.S.Shellard and S.J.Rankin (Cambridge University
Press, Cambridge) 693 (2003).\par
\indent
[26] P.D.D'Eath, {\it Black Holes: Gravitational Interactions}, (Oxford
University Press, Oxford) (1996).\par
\indent
[27] G. 't Hooft, {\it Phys. Lett. B} {\bf 198}, 61 (1987).\par
\indent
[28] S. Giddings, 'Black holes at accelerators', in {\it The Future of
Theoretical Physics and Cosmology}, eds. G.W.Gibbons, E.P.S.Shellard
and S.J.Rankin (Cambridge University Press, Cambridge) 278 (2003).\par
\indent
[29] J.Wess and J.Bagger, {\it Supersymmetry and Supergravity},
2nd. edition, (Princeton University Press, Princeton) (1992).\par
\indent
[30] P.D.D'Eath, 'Loop amplitudes in supergravity by canonical
quantization', in {\it Fundamental Problems in Classical, Quantum and
String Gravity}, ed. N.S{\'a}nchez (Observatoire de Paris) 166
(1999), hep-th/9807028.\par
\indent
[31] M.B.Green, J.H.Schwarz and E.Witten, {\it Superstring Theory,
vols. 1 and 2}, (Cambridge University Press, Cambridge) (1987).\par
\indent
[32] M.B.Green, 'A brief description of string theory' in {\it The
Future of Theoretical Physics and Cosmology} eds. G.W.Gibbons,
E.P.S. Shellard and S.J.Rankin (Cambridge University Press, Cambridge)
473 (2003).\par
\indent
[33] T.Regge and J.A.Wheeler, {\it Phys. Rev.} {\bf 108}, 1063 (1957).\par 
\indent
[34] J.Mathews, {\it J. Soc. Ind. Appl. Math.} {\bf 10}, 768 (1962).\par
\indent
[35] J.N.Goldberg, A.J.MacFarlane, E.T.Newman, F.Rohrlich and 
E.C.G.Sudarshan,\par
\noindent
{\it J. Math. Phys.} {\bf 8}, 2155 (1967).\par
\indent
[36] C.V.Vishveshwara, {\it Phys. Rev. D} {\bf 1}, 2870 (1970).\par
\indent
[37] F.J.Zerilli, {\it Phys. Rev. D} {\bf 2}, 2141 (1970).\par
\indent
[38] W.L.Burke, {\it J. Math. Phys.}  {\bf 12}, 401 (1971).\par
\indent
[39] J.A.H.Futterman, F.A.Handler and R.A.Matzner,   
{\it Scattering from Black Holes},\par
\noindent 
(Cambridge University Press, Cambridge) (1988).\par 
\indent
[40] M.W.Choptuik, 'Critical Behaviour in Massless Scalar-Field
Collapse', in {\it Approaches to Numerical Relativity},
ed. R.d'Inverno, (Cambridge University Press, Cambridge) (1992).\par
\indent
[41] P.D.D'Eath and A.Sornborger, {\it Class. Quantum Grav.} {\bf 15},
3435 (1998).\par
\indent
[42] D.Brill and J.B.Hartle, {\it Phys. Rev.} {\bf 135}, 1327 (1964).\par
\indent
[43] R.Isaacson, {\it Phys. Rev.} {\bf 166}, 1263, 1272 (1968).\par
\indent
[44] J.D.Jackson, {\it Classical Electrodynamics}, (Wiley, New York) 
(1975).\par
\indent
[45] A.N.St.J.Farley and P.D.D'Eath, 'Quantum Amplitudes in Black-Hole
Evaporation', paper being submitted.\par
\indent
[46] M.Abramowitz and I.A.Stegun, {\it Handbook of Mathematical
Functions}, (Dover, New York) (1964).\par
\indent
[47] N.D.Birrell and P.C.W.Davies, {\it Quantum fields in curved
space}, (Cambridge University Press, Cambridge) (1982).\par
\indent
[48] V.P.Frolov and I.D.Novikov, {\it Black Hole Physics}, (Kluwer
Academic, Dordrecht) (1998).\par
\indent
\bigskip

\bye